\documentclass[french]{article-hermes}
\usepackage[frenchb]{babel}
\usepackage[T1]{fontenc}

\ifx\pdftexversion\undefined
  \usepackage[dvips]{graphicx}
  \DeclareGraphicsExtensions{.eps} 
\else
  \usepackage[pdftex]{graphicx}
  \DeclareGraphicsExtensions{.jpg,.mps,.pdf,.png} 
\fi

\title[Flexibilité dans OpenCCM]{Reconfigurations dynamiques de services dans
  un intergiciel à composants CORBA CCM}

\author{Assia Hachichi, Cyril Martin, Ga\"el Thomas, Simon Patarin, Bertil Folliot}
\submitted{Deploiement et (Re) Configuration de Logiciels}{1}

\address{%
Laboratoire d'Informatique de Paris 6 (lip6)\\
8 rue du Capitaine Scott\\
75015 Paris\\
\{assia.hachichi,gael.thomas,cyril.martin,bertil.folliot\}@lip6.fr, simon.patarin@inria.fr}

\abstract{
Today, component oriented middlewares are used to design, develop and deploy
easily distributed applications, by ensuring the heterogeneity,
interoperability, and reuse of the software modules, and the separation between
the business code encapsulated in the components and the system code managed by
the containers. 
Several standards answer this definition such as: CCM (CORBA Component Model),
EJB (Enterprise Java Beans) and .Net. However these standards offer a limited
and fixed number of system services, removing any possibility to add system
services or to reconfigure dynamically the middleware. 
Our works propose mechanisms to  add and to adapt dynamically the system
services, based on a reconfiguration language which is dynamically  adaptable
to the need of the reconfiguration, and on a tool of dynamic reconfiguration, a
prototype was achieved for the OpenCCM platform, that is an implementation of
the CCM specification. This work was partially financed by the european project
IST-COACH (2001-34445).
}

\resume{
De  nos jours, les intergiciels à composants sont utilisés pour concevoir, 
développer, et déployer facilement les applications réparties, et assurer
l'hétérogénéité, et l'interopérabilité, ainsi que la réutilisation des modules
logiciels, et la séparation  entre le code métier encapsulé dans des composants
et le code système géré par les conteneurs. 
De nombreux standards répondent à cette définition tels: CCM (CORBA Component
Model), EJB (Entreprise Java Beans) et .NET. Cependant ces standards offrent un
nombre limité et figé de services systèmes, supprimant ainsi toute possibilité
d'ajout de services systèmes ou de reconfiguration dynamiques de l'intergiciel. 
Nos travaux proposent des mécanismes d'ajout et d'adaptation dynamique des
services systèmes, basés sur un langage de reconfiguration adaptable
dynamiquement aux besoins de la reconfiguration et sur un outil de
reconfiguration dynamique. Un prototype a été réalisé pour la plateforme
OpenCCM, qui est une implémentation de la spécification CCM de l'OMG. Ce travail a été  partiellement financé par le projet européen IST-COACH (2001-34445).
}
\motscles{Adaptation et reconfiguration dynamique, conteneurs adaptables, CCM, MVV}

\proceedings{DECOR'04, Déploiement et (Re)Configuration de Logiciels}{159}

\begin{document}

\maketitlepage

\section{Introduction}
Les systèmes informatiques sont de plus en plus complexes et difficiles à maintenir et les
différents éléments constitutifs d'un environnement sont souvent physiquement répartis. Les intergiciels
ont été introduits pour résoudre ces difficultés et pour 
proposer des mécanismes systèmes génériques communs à la plupart des applications réparties,
comme la persistance, les transactions ou le nommage.

La dernière génération d'intergiciels à base de composant introduit la notion de conteneur. Un conteneur
permet de découpler le code métier du code système en encapsulant le code métier et en gérant les
services systèmes de manière transparente pour l'application.
Le développement d'applications basées sur des composants est aujourd'hui
largement utilisé, mais ces plateformes sont encore inadaptées aux besoins
spécifiques en terme d'adaptation de services systèmes~: les conteneurs n'offrent pas de
moyen de reconfigurer, d'ajouter ou de supprimer dynamiquement des services
systèmes. 
L'adaptation de l'application nécessite ainsi de modifier un ou plusieurs
code(s) source(s), de recompiler et de redémarrer l'application en cours.

Dans ce contexte, notre travail propose des mécanismes permettant l'ajout
et la gestion des services systèmes dynamiquement, sans
avoir à modifier ni le code source des composants sur lesquels les
services s'appliquent, ni le code source de l'intergiciel ciblé. Notre approche
est guidée par la notion de \emph{réutilisabilité}~: le but n'est pas de créer un nouvel intergiciel
mais bien de proposer une solution pour adapter dynamiquement des applications déjà
écrites. Les mécanismes d'adaptation
sont basés sur un langage de reconfiguration adaptable dynamiquement, et sur un
outil d'adaptation dynamique appelé CVM~(Container Virtual Machine). 

Un premier prototype a été réalisé sur la plateforme OpenCCM~\cite{openccm}, qui est une
plateforme logicielle ouverte implémentant la spécification des composants
CCM définie par le consortium Object Management Group (OMG). Ce prototype
permet l'ajout et la reconfiguration de nouveaux services systèmes, et offre la possibilité à un administrateur de
spécifier et de déployer dynamiquement des propriétés systèmes non prévues
initialement.

Lors de ces travaux, nous nous sommes fixé comme objectifs d'offrir des
mécanismes permettant de gérer les services systèmes dans les plateformes
actuelles, sans modifier les codes sources, de façon dynamique et transparente pour l'application. La
conception de ces mécanismes demande la prise en compte de quatre caractéristiques
importantes~:

\begin{itemize}
\item définir un mécanisme permettant d'intégrer des services systèmes dans les conteneurs CCM,
\item offrir un langage de reconfiguration extensible dynamiquement et générique par rapport à l'intergiciel ciblé,
\item définir une architecture permettant de reconfigurer dynamiquement des intergiciels, en
intégrant de nouveaux services systèmes dynamiquement,
\item offrir un moyen d'administrer la reconfiguration dynamique des services à distance.
\end{itemize}

Dans ce qui suit, la section~\ref{stat} présente d'autres propositions permettant de rendre
des intergiciels flexibles. 
Ensuite, notre proposition d'adaptation dynamique est
détaillée dans la section~\ref{archi}. La section~\ref{evaluation} décrit deux
exemples d'adaptation dynamique, la section~\ref{performances} présente les
performances de la solution proposée, et la section~\ref{perspectives} 
présente les conclusions et perspectives du projet.

\section{Travaux similaires}\label{stat}

Plusieurs modèles à composant existent tels que : Microsoft .Net, les
Entreprise Java Beans de Sun Microsystem ou encore CORBA Component Model de
l'OMG. Ces modèles sont de plus en plus utilisés pour concevoir et déployer des
applications réparties. Cependant, ils n'offrent pas la possibilité d'ajouter
ou de reconfigurer un service système après le déploiement initial de l'application.

Les premiers intergiciels n'ont pas été conçus pour être flexibles. 
Toutefois, des possiblités d'adaptation ont été proposées 
tels que les intercepteurs, et le Portable Object Adaptor (POA) dans CORBA~(\cite{coeur}, chapitres 7,8 et 9). 
Les intercepteurs~\cite{intercepteurs-corba} permettent d'inserer du code avant la
réception et après l'envoie d'une requête.
Le POA permet de contrôler finement la politique de l'adaptateur d'objet.

Certains travaux visent à rendre CORBA plus flexible.
DynamicTAO~\cite{dynamicTAO, 2K} (basé sur TAO~\cite{tao}) est 
un environnement CORBA réflexif.
DynamicTAO réifie des éléments internes de l'ORB sous la forme de composants
appelés composants de configuration. 
Deux exemples d'adaptation dynamiques sont présentés dans~\cite{dynamicTAO}~: 
l'ajout d'un service de monitoring et l'ajout d'un service de sécurité. 
DynamicTAO permet de garder une compatibilité avec les applications CORBA 
tout en offrant un haut degré d'adaptabilité.
Une des difficultés que soulève ce projet est 
le problème de la cohérence lorsqu'on remplace une politique par une autre~: 
l'exemple donné est le remplacement d'une gestion de thread. 
Pour passer d'une politique de pool de thread à une autre politique, 
il est nécessaire que le pool soit vide.
Cette information doit être transmise par l'ancienne politique, elle doit donc
avoir été conçue de manière à pouvoir être remplacée.

AspectIX~\cite{hauck98aspectix} présente une architecture d'intergiciel 
basée sur le modèle d'objets à fragments~\cite{makpangou94fragmented}.
Les fragments peuvent masquer la réplication d'un objet distribué, imposer des contraintes temps
réelles au canal de communication, mettre en cache des données de l'objet etc.
Ces aspects non fonctionnels peuvent être configurés 
via une interface générique de l'objet...
Chaque objet global peut être configuré par un profil 
qui spécifie les aspects que doivent respecter les fragments.
Quatre profils sont planifiés, en particulier un profil CORBA 
qui permet de faire interagir les objects AspectIX avec CORBA.
Cette approche permet de séparer l'application de l'intergiciel dans lequel elle est déployée.

Une architecture de conteneurs ouverts est proposée dans~\cite{merle}. 
Cette architecture permet d'adapter et d'étendre dynamiquement les
fonctions systèmes,
et elle permet d'exposer un certain nombre de propriétés du conteneur 
grâce à des mécanismes d'interception, de coordination 
(pour ordonner les appels de fonctions systèmes) et de contrôle.

Des travaux comme Oopp~\cite{oopp} introduisent un système d'interception 
et de redéfinition de l'appel d'opération dans un ORB.

JavaPOD~\cite{javapod}, un modèle de composant proche des EJB, offre 
la possibilité d'attacher des propriétés non fonctionnelles aux composants 
grâce à des conteneurs ouverts et extensibles.

Ces différents travaux augmentent les possibilités d'un intergiciel en le
recodant. Le direction prise par nos travaux est assez différente puisque nous
utilisons les propriétés actuelles des intergiciels pour permettre à un
administrateur de modifier dynamiquement les propriétés de l'application et de
l'intergiciel. Les intergiciels offrent la possibilité de déployer une
application, nous utilisons les mêmes mécanismes pour redéployer dynamiquement les composants
systèmes de l'application.

\section{Architecture}\label{archi}

Actuellement, les plateformes des intergiciels existantes 
sont monolithiques ce qui empêche toute adaptation dynamique 
des mécanismes internes, et des systèmes sous-jacent. 
Pour cela nous avons conçu un outil assurant la reconfiguration 
à la volée, sans modifier ou réécrire ces plateformes. 
L'idée principale de la CVM est d'ajouter au déploiement initial, 
de façon transparente, un point d'entrée qui peut interagir 
sur la plateforme et l'application ciblée.

L'architecture de cet outil permet d'agir dynamiquement sur la plateforme
en manipulant ses symboles et ses méthodes, permettant ainsi d'ajouter
et reconfigurer des services systèmes non prévus initiallement. 

\subsection{Définition d'un langage de reconfiguration}
L'adaptation dynamique nécessite la définition d'un langage 
et une implémentation qui étendent les opérations d'accès usuels 
pour contrôler ce qui peut être adapté. 
La définition d'un langage de reconfiguration permet de séparer 
la logique de reconfiguration de sa mise en oeuvre.
Ce langage doit être 
à la fois adaptable dynamiquement aux nouveaux besoins de reconfiguration, 
et générique par rapport à l'intergiciel ciblé dans un but de réutilisabilité.
Ceci permet de réduire les possibilités de reconfiguration en
limitant les symboles du langage, 
ou bien au contraire d'étendre le langage en autorisant 
et l'instrospection de l'environnement et la création de nouveaux symboles.

La plateforme de reconfiguraton choisie est la Machine
Virtuelle Virtuelle~\cite{ogel03, folliot98}~(MVV), une plateforme hautement adaptable
permettant de construire son environnement d'exécution et son langage dédié dynamiquement.
La MVV permet à la fois de modifier les mécanismes mis en oeuvre pour reconfigurer l'environnement
mais aussi d'étendre ou modifier le langage associé.

\subsection{Mécanisme de reconfiguration à distance}
Afin de permettre une administration distante, nous avons construit 
un environnement de reconfiguration à distance pour la MVV.
Une reconfiguration initiale doit être chargée sur toutes les MVV.

\begin{figure}[ht]\centering
\includegraphics[width=12cm]{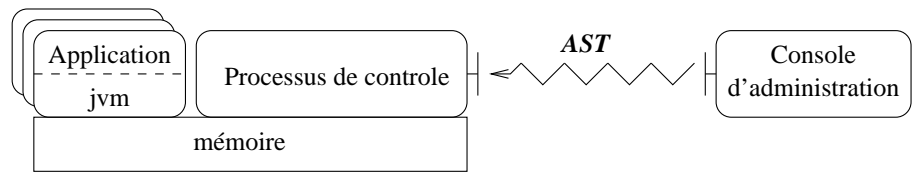}
\caption{Mécanisme de reconfiguration à distance\label{DVM}}
\end{figure}

La MVV est séparé en deux entités distinctes~: la première s'occupe de
parser/lexer des scripts qui reconfigurent l'environnement cible, 
ces scripts sont transformés en arbre de syntaxe abstraite et envoyés à la
seconde (Figure~\ref{DVM}).
De l'autre côté, la MVV a été modifiée de manière à recevoir les arbres sur un
canal de communication (exemple: une socket TCP). L'arbre
est compilé et exécuté sur la machine distante.

L'administration distante se fait en deux phases~:
\begin{enumerate}
\item {{\bf Déploiement initial : }
Toutes les MVVs chargent le script de reconfiguration et se mettent en attente de connexion.}
\item {{\bf Reconfiguration : }
Elle suit le modèle client/serveur, les machines serveurs attendent 
d'être reconfigurées et le client est la console d'administration.
Le client ouvre un canal de communication vers un serveur et lui envoie une séquence d'instructions.}
\end{enumerate}

Dans le cas de la CVM, les processus serveurs s'exécutent dans le même environnement qu'OpenCCM.
La machine cliente permet donc d'administrer à distance l'environnement OpenCCM.

\subsection{Intégration de service}
Deux méthodologies d'intégration de service ont été proposées, leurs
réalisations sont présentées ci-dessous :

\begin{itemize}
\item {{\bf  Technologie des intercepteurs :} La spécification
  Corba~\cite{Obj02a} définit l'interface de l'intercepteur
  portable~\cite{Obj02b}, pour insérer des crochets directement dans l'ORB. Ces
  crochets sont actifs pour chaque opération effectuée. 
  Les crochets peuvent être localisés soit du côté client soit du côté
  serveur. En particulier, il est possible de connaître l'identificateur
  unique de la transaction, le nom exact de la méthode invoquée, le nom de
  l'expéditeur de la requête, et le statut de la réponse. 
  Il est aussi possible d'associer une donnée privée, qui est piggybacked, à
  travers chaque invocation de crochet dans la même requête. Ceci permet, par
  exemple, d'associer un timestamp à l'étape du traitement de la requête.

Naturellement, le passage de toutes les requêtes par la couche d'intercepteur 
a un surcoût particulier. 
Des travaux \cite{MVB01} montrent que l'activation des intercepteurs portables
augmente la latence par un facteur variant entre 2\% et 10\%, 
et le temps de taitement des requêtes par un facteur entre 1,5\% et 16\%. 
Ces  résultats démontrent que cette technologie à un impact réduit 
sur l'exécution global des applications. 
Ce coût est relativement limité, ce qui nous a motivé à utiliser
cette technique pour déployer et reconfigurer les services systèmes 
en cours d'exécution.

}

\item {{\bf Composant Orienté Systèmes}
La réalisation de cette méthodologie consiste à récupérer la référence de
l'ORB et du conteneur à adapter, puis à ajouter un composant CCM dont le code
est celui du service système, et à effectuer les connexions nécessaires avec
les composants sur lesquels ce service va s'appliquer. Le point fort de cette
méthodologie est qu'elle est générique, elle peut s'appliquer pour tous les
intergiciels à composants. 
Les méthodes d'intégration présentées seront utilisées par la suite.

}
\end{itemize}

\subsection{Réalisation de la CVM}
L'idée principale de l'architecture CVM est d'ajouter un point d'entrée dans
l'application, au déploiement initial.
Dans le cas d'OpenCCM, la réalisation de ce point d'entrée se fait à l'aide
d'une méthode native Java, qui lance la MVV.
L'interfaçage entre la MVV et la JVM standard est réalisée 
par JNI~(Java Native Interface~\cite{jni}). 
JNI est une interface entre les fonctions natives et la machine virtuelle Java. 
La MVV est exécutée par un thread Java en concurrence avec ceux de l'application.  
Le langage de la MVV est ensuite enrichi dynamiquement par les fonctions JNI~: 
les scripts écrits pour la MVV peuvent alors interagir directement avec la JVM, 
et la MVV est capable de manipuler les méthodes et les symboles de l'application Java.

Une reconfiguration est constitué de deux phases importantes :
\begin{enumerate}
\item La première phase consiste à construire dans la MVV les mots clés du langage
  de reconfiguration dynamique d'OpenCCM. 
  Ces mots clés permettent de récupérer la réference de l'ORB, 
  d'ajouter ou de supprimer des composants, 
  de charger des classes, ou bien d'invoquer des méthodes. 
\item Le deuxième phase consiste à écrire un script MVV contenant 
  les adaptations souhaitées. 
  Ce script est chargé à distance dans la console d'administration et executé
  dans les CVM. 
  Les scripts peuvent soit étendre le langage de reconfiguration, soit utiliser
  les mots clés déjà construits pour modifier l'application OpenCCM.
\end{enumerate}

Deux exemples de reconfigurations sont illustrés dans ce qui suit.
\begin{figure}[ht]\centering
\includegraphics[width=12cm]{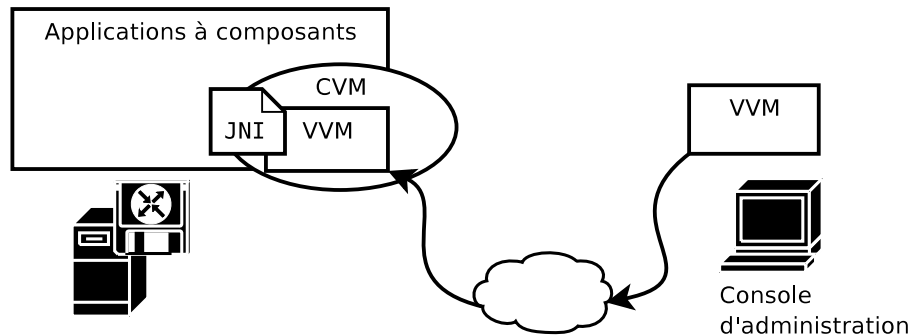}
\caption{Fonctionnement de la CVM  (Container Virtual Machine)}
\end{figure}

\section{Exemples d'utilisation de la CVM}\label{evaluation}
Dans cette section, deux exemples d'utilisation de la CVM sont présentés. 
Le premier consiste à ajouter un service de monitoring flexible, basé sur la
technique d'intercepteur. 
Le second consiste à ajouter un service de cryptage, basé sur la technique 
de COS (composant orientée systèmes), puis de l'adapter dynamiquement.

\subsection{Service de monitoring flexible}
Les services de monitoring sont la base de la gestion des applications
distribuées complexes. 
Ils permettent, par exemple, d'avoir des informations sur les
ressources utilisées et de définir des statistiques et des profils
utilisateurs.  Ces informations sont nécessaires pour adapter le logiciel
au nouveau contexte de l'environnement. 
Dans le  cas des applications CCM, qui s'exécutent pendant une longue
durée et dans un environnement à large échelle, 
il est souvent utile de reparamétrer un composant spécifique 
ou de remplacer l'implémentation d'un composant par un équivalent. 

L'objectif de ce service de monitoring est de pouvoir collecter des
statistiques sur la façon dont les composants interagissent, et de rendre
ces informations disponibles à un <<service de reconfiguration>> capable
de les utiliser pour adapter la plate-forme. 
En particulier, cela signifie que l'on s'intéresse exclusivement 
à la récolte et au traîtement statistique des données 
et non au schéma intelligent qui saurait comment les
interpréter et réagir en conséquence. Comme on ne veut pas catégoriser
l'application monitorée ou le logiciel spécifique (ORB, plate-forme CCM, ...)
utilisé, le code source de ceux-ci ne doit pas être modifié.  Cette hypothèse
est importante car beaucoup d'ORB sont des programmes commerciaux dont le
code source n'est pas toujours disponible. En outre, vu la diversité des
applications et des environnements visés, le service doit être flexible. Cette
flexibilité doit permettre aux programmeurs d'applications et aux
administrateurs systèmes de définir leurs propres paramètres (ou d'adapter des
paramètres existants) et la façon dont les données sont collectées, en fonction
de leurs besoins.
Nous présentons dans ce qui suit les métriques utilisées pour implémenter ce
service de monitoring flexible, puis son implémentation dans la plateforme
OpenCCM et son intégration dynamique dans l'application en cours d'exécution.

\subsubsection{Métriques} 
Plusieurs métriques peuvent être extraites des données brutes
obtenues par les intercepteurs portables. Les métriques
disponibles sont présentées dans ce qui suit:

\begin{enumerate}

\item{ {\bf Métriques comptables :}
Ce type (simple) de métrique est basé sur les compteurs qui tracent les occurrences des opérations similaires.
\begin{itemize}
\item {{\bf Invocations de méthode :} compte le nombre de fois où une méthode spécifique est appelée. }
\item  {{\bf utilisation composant :} additionne toutes les invocations des méthodes appartenant au même composant.}
\end{itemize}
}

\item{ {\bf Métriques temporelles :}
  Associées avec des timestamps précis, qui sont fournis par la plupart des systèmes d'exploitation modernes, les traces produites peuvent être
employées pour extraire la métrique temporelle, mesurant le temps requis pour
  effectuer une opération spécifique. 

}

\end{enumerate}

\subsubsection{Implémentation du service de monitoring dans la plateforme OpenCCM}
Les intercepteurs portables ont déjà été implémentés dans la plateforme
OpenCCM. En effet  les développeurs d'OpenCCM les utilisent pour
fournir le service de trace. Ce service est activé en ajoutant $ --trace $ à
n'importe quel programme déployé.
Ce service de trace nous aide de deux façons dans l'implémentation du service
de monitoring. Premièrement il montre précisément les informations disponibles
aux intercepteurs portables dans la plateforme OpenCCM. Deuxièmement, il
simplifie et propose une méthodologie à suivre pour initialiser les
intercepteurs portables avec OpenCCM.
Les lignes suivantes montrent un exemple des enregistrements produits
par le service~:

{\footnotesize
\begin{verbatim}
     INFO date=2003-09-03 17:48:06,607 thread=Thread pool thread \#1
     topic=org.omg.openccm.core.CORBA.interceptors.server
     class=org.objectweb.corba.trace.PI.TraceSI
     method=send\_reply line=93 request id = 52 operation = create
     arguments = exceptions = response expected = true
     reply status = SUCCESSFUL
     target most derived interface = IDL:DiningPhilosophers/ForkHome:1.0
\end{verbatim}
}

Ce service enregistre dans un fichier journal toutes les informations sur les
requêtes invoquées, scrute périodiquement ce journal, puis fournit les
statistiques sur les appels effectués.

\subsubsection{Intégration dynamique du service de monitoring flexible }
L'utilisation de la CVM pour intégrer dynamiquement le service de monitoring,
est consituée de deux étapes: 

\begin{enumerate}
\item Spécifer et ajouter les opérations d'adaptation dans la MVV, tel que
  l'ajout d'url dans le class loader d'OpenCCM, 
  la récupération de la réference de l'ORB, 
  ou l'invocation des méthodes Java chargées par le class loader d'OpenCCM.
\item Écrire et charger le script MVV permettant d'intégrer le service.
\end{enumerate}
 En exemple le script VVM ajoutant le service de monitoring  dynamiquement :
\begin{small}
\begin{verbatim}
(add_url_classloader "file:/home/user/openccm/Monitor/")
(add_url_classloader "file:/home/user/ORB/OpenCCM_Plugins.jar")

(define clazz "Monitor")
(define sign "(Ljava/lang/Object;)V")
(define mon (runCCM clazz "getInstance" sign))

(define sign  "(Ljava/lang/Object;)Ljava/lang/String;")
(define logfile "/tmp/monitor.log")
(define metric (runCCM "DebugMetric" "DebugMetric" sign log))

(define sign  "(Ljava/lang/Object;)Ljava/lang/Object;")
(define handle (runCCM_arg clazz "registerMetric" sign mon metric))

(runCCM_arg clazz "start" "()V")
\end{verbatim}
\end{small}
\subsection{Service de cryptage}
Soit une application contenant deux composants A et B, qui se trouvent
respectivement dans les conteneurs CA et CB, où A envoie régulièrement des
messages à B. Considérons que durant l'exécution, l'administrateur décide
d'envoyer des messages cryptés à B. Pour ce faire, l'intégration d'un service
de cryptage revient à ajouter un composant COS de cryptage en utilisant la
CVM, dans le conteneur CA, puis d'établir les connections nécessaires.
\begin{figure}[ht]\centering
\includegraphics[width=12cm]{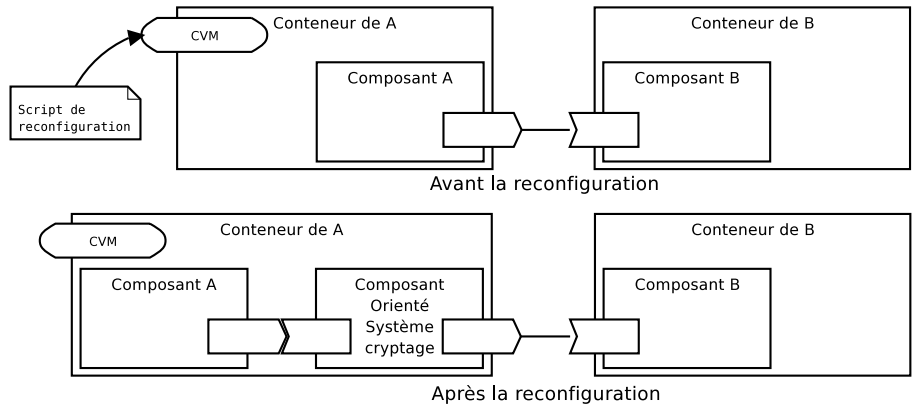}
\caption{Intégration du service de cryptage}
\end{figure}

\subsubsection{Implémentation du COS de cryptage}
L'implémentation du COS passe par deux étapes, qui sont:
\begin{enumerate}
\item {Définir des opérations
  nécessaires à ajouter dans la MVV pour intégrer le COS de cryptage, tels que:

\begin{itemize}
\item {{\bf Récupération de la référence de l'ORB :} $(clssLoaderCCM getorb  )$
Cette instruction fait appel à une méthode Java en utilisant la référence de
l'environnement JNI. }
\item {{\bf Chargement d'une classe} : $(jrun~"nameClass")$ 
Cette instruction nous permet de charger dans la JVM de l'application OpenCCM
la classe « nameClass ». Ce chargement peut se faire soit dans le classLoader
par défaut ou dans le ClassLoader d'OpenCCM.}
\end{itemize}
\item {Écrire puis charger le script MVV qui
  permet d'ajouter le COS dans le conteneur CA, de déconnecter A et B, et
  d'établir la connexion de A vers COS et de COS vers B. 
}
}
\end{enumerate}

\subsubsection{Adaptation du service de cryptage}
Adapter le comportament d'un composant peut être réalisé en remplaçant le
composant par un nouveau.
Cependant il est plus simple et moins coûteux d'adapter le composant en remplaçant
certaines de ses méthodes.

Dans le cas de l'adaptation du COS de cryptage, 
il suffit d'adapter la méthode Java qui implémente l'algorithme de cryptage. 
Le standard Java permet le chargement dynamique de classe 
et la surcharge des méthodes de sérialisation. 
En couplant la plateforme Java de Sun et la  CVM nous pouvons adapter 
une méthode Java. 
Prenons l'exemple de la méthode « metA » de la classe A, 
l'adaptation de cette classe se fait en chargeant une nouvelle classe A1 
qui hérite de A, et qui implémente le nouveau code de la méthode « metA », 
puis en redirigeant tous les appels vers la nouvelle méthode chargée. 

Cette méthodologie  permet d'adapter une méthode statique Java, 
Cependant elle augmente le nombre de classes chargées en mémoire, 
car la JVM standard ne prévoit pas le déchargement des classes.

\section{Mesures de performances}\label{performances}
Pour les deux exemples cités précédemment, nous avons mesuré la moyenne de la
durée de la reconfiguration sur un pentium III candensé à 664MHz sous Linux. 
La moyenne de la durée d'intégration du service de monitoring, qui est basé sur les
intercepteurs portables, est de 8,539 secondes. La moyenne de la durée d'ajout du COS de
cryptage est de 2,054 secondes. L'intégration du COS est légèrement plus
rapide que celle des intercepteurs portables. Cependant ce coût reste limité et
comme les serveurs restent actifs  pendant cette periode, il n'y a pas de
rupture du service assuré.

\section{Conclusion}\label{perspectives}
Dans cet article, nous fournissons une réponse à la problématique de
l'adaptation des services systèmes dans les plateformes actuelles 
qui sont fermées. 
D'abord, nous proposons deux méthodologies pour intégrer 
les services systèmes ; l'une basée sur la technologie des intercepteurs
portables, et l'autre sur des composants CCM orientés système. 
Ensuite nous  définissons l'architecture d'un outil (appelé 
Container Virtual Machine) permettant  de reconfigurer dynamiquement 
les services systèmes, sans avoir pour autant à modifier ni le code 
des composants sur lesquels le service s'applique ni le code de l'intergiciel. 
Les modifications se font dynamiquement, sans interruption de service.
Finalement, nous avons présenté la réalisation du premier prototype
pour la plateforme OpenCCM du LIFL.

L'outil CVM est basé sur une architecture généraliste par rapport
à l'intergiciel ciblé, il offre un moyen de contrôler et de redéfinir
dynamiquement les opérations d'adaptation à autoriser, mais il permet aussi
l'intégration dynamique des services systèmes. Ces services peuvent être écrits
soit en Java, ou bien en MVV ce qui permettra de faire appel à des opérations de bas
niveau. Nous avons présenté  ces adaptations à l'aide de deux exemples~:
l'ajout dynamique du service de  monitoring et l'intégration
dynamique d'un composant de cryptage. 

En  conclusion, la CVM offre la possibilité de spécifier les propriétés
systèmes et de les intégrer dynamiquement. 
Ses spécifications sont valides pour la norme CORBA CCM, 
et se basent sur un langage de reconfiguration généraliste, 
indépendant de l'intergiciel ciblé.

En perspective, nous visons à généraliser notre méthodologie d'adaptation à
d'autres intergiciels à composants.

\bibliography{main}

\end{document}